\documentclass[12pt]{article}
\usepackage{graphics}

\def\<~{\stackrel{<}{\mbox{\scriptsize $\sim$}}}
\def\>~{\stackrel{>}{\mbox{\scriptsize $\sim$}}}
\def\eq#1{(\ref{#1})}

\begin{document}

\title{
Unstable Axion Quintessence Revisited}
\author{
Carl L. Gardner\\
{\em gardner@math.asu.edu}\\
School of Mathematical \& Statistical Sciences\\ 
Arizona State University\\
Tempe AZ 85287-1804
}
\date{}

\maketitle
\thispagestyle{empty}

\begin{abstract}

In axion quintessence, the cosmological era with an energy contrast in
dark energy $0.1 \le \Omega_{DE} \le 0.9$ may represent a significant
fraction of the universe's lifetime if the minimum of the axion
potential is negative (unstable axion quintessence), thus resolving
the cosmic coincidence problem, as pointed out by Kallosh, Linde,
Prokushkin, and Shmakova.  Further details of the evolution of the
quintessence field, the scale factor of the universe, and the Hubble
parameter are presented here, focussing on models with $\Omega_{DE,0}$
= 0.72 and recent dark energy average equation of state parameter $-1
< w_0 < -0.85$.  For these parameter values, the contracting universe
enters a late time era of kination, the negative Hubble parameter
acting like a negative friction term in the Klein-Gordon equation, and
the axion field makes many transits of---but never remains in---its
vacuum state.

Robust, scaled cosmological equations are derived for simulating the
evolution of the scalar field, the scale factor, and the Hubble
parameter during a contracting phase of the universe.  These equations
allow the simulations presented here to proceed much closer to the
singularity at the end of the collapsing universe than any previous
simulations.

\end{abstract}

\section{Introduction}

Typically in quintessence theories with an asymptotically vanishing
effective cosmological constant, the energy contrast in dark energy
$\Omega_{DE}$ rises from near zero for redshifts $z > 5$ to near one
for $z < -0.5$, mimicking a true cosmological constant.  At late times
the quintessence field may begin to oscillate about its minimum,
behaving like nonrelativistic matter, or the quintessence field may
evolve toward infinity---in both cases with vanishing vacuum energy.
In such theories, there is a period between roughly 3.5 Gyr and 20 Gyr
after the big bang when $0.1 \le \Omega_{DE} \le 0.9$.  However if the
universe continues to expand forever, or even if positive curvature
begins to dominate at late times (after the quintessence field has
evolved to its minimum) and the universe enters a contracting stage,
this period when the energy densities of dark energy and matter are
comparable is a small or vanishing fraction of the total lifetime of
the universe.  This is called the cosmic coincidence problem.

However in axion quintessence (as in other unstable de Sitter
quintessence models), the cosmological era with $0.1 \le \Omega_{DE}
\le 0.9$ may represent a significant fraction of the universe's
lifetime if the minimum of the axion potential is negative (unstable
[de Sitter] axion quintessence), thus resolving~\cite{Kallosh:2002gf}
the cosmic coincidence problem.  (Negative $\rho_\Lambda$ or
$\rho_{DE}$ and the fate of the universe are discussed in
Refs.~\cite{Krauss:1999br} and~\cite{Kallosh:2002gf} plus references
therein.)

Vilenkin~\cite{Vilenkin:2009zz} summarizes the currently predominant
view that the dark energy is a cosmological constant, and points out
the major weaknesses of most quintessence models: (i) most
quintessence models assume that $\rho_{DE} \rightarrow 0$ as the
quintessence field $\phi$ evolves toward its minimum (which may be at
$|\phi| \rightarrow \infty$), conflicting with the expectation that
$\rho_{DE}$ ought to evolve to a nonzero vacuum energy density; (ii)
most quintessence models do not solve the coincidence problem that
$\rho_{DE,0} = \rho_\Lambda \sim \rho_{m0}$ (the subscript ``0'' will
denote present values); and (iii) since the present dark energy
average equation of state parameter $w_0 \approx -1$, perhaps $w_0
\equiv -1$ and quintessence models are irrelevant.

The unstable axion quintessence potential $V(\varphi) = A
\cos(\varphi)$, where $\varphi \equiv \phi/M_P$ and the Planck mass
$M_P = 1/\sqrt{8 \pi G} = 2.4 \times 10^{18}$ GeV, addresses all of
these issues, since the facts that the minimum of the potential is at
$-A \approx -\rho_\Lambda$, $\rho_{DE,0} = \rho_\Lambda$, and $w_0 \ne
-1$ but $\approx -1$ are interrelated aspects of the model, and occur
for an appreciable range of initial values for $\phi$.

For $V(\varphi) = A \cos(\varphi)$, the initial value of the scalar
field need only satisfy $0 \le \varphi_i/\pi \le 0.23$ to produce a
universe like ours~\cite{Gardner:2004in} (due to symmetry, we can
restrict our attention to $0 \le \varphi_i \le \pi$).  Thus there is a
significant 23\% range of the possible initial values $\varphi_i$
which will produce a universe like ours.\footnote{Qualitatively
  similar results to those presented here are obtained for $V(\varphi)
  = A \cos(\lambda \varphi)$ for $\lambda = O(1)$.}  For these initial
values, the contracting universe enters a late time era of kination
(during which the scalar field kinetic energy dominates over all other
forms of energy), the negative Hubble parameter acting like a negative
friction term in the Klein-Gordon equation, and the axion field makes
many transits of---but never remains in---its vacuum
state.\footnote{The coupling of the quintessence field to other
  particles must be very small, and will for the most part be
  neglected in this investigation.}

In Section 2, the basic cosmological equations are presented for the
evolution of the scalar field, the scale factor, and the Hubble
parameter, and cast in the form of a scaled, dimensionless system of
first-order equations in the conformal time, appropriate for a
contracting (or expanding) universe.  These equations allow the
simulations presented in Section 3 (see
Figs.~\ref{fig-V}--\ref{fig-KE-V}) to proceed much closer to the
singularity at the end of the collapsing universe than any other
simulations presented in the literature, and provide the basis for a
more detailed analysis of the last stages of the collapsing universe
than has appeared before.

\section{Cosmological Equations}

In the quintessence/cold dark matter (QCDM) model, the total energy
density $\rho = \rho_m + \rho_r + \rho_\phi$, where $\rho_m$,
$\rho_r$, and $\rho_\phi$ are the energy densities in
(nonrelativistic) matter, radiation, and the axion quintessence scalar
field $\phi$, respectively.  Ratios of energy densities to the
critical energy density $\rho_c$ for a flat universe will be denoted
by $\Omega_m = \rho_m/\rho_c$, $\Omega_r = \rho_r/\rho_c$, and
$\Omega_\phi = \rho_\phi/\rho_c$, while ratios of present energy
densities $\rho_{m0}$, $\rho_{r0}$, and $\rho_{\phi0}$ to the present
critical energy density $\rho_{c0}$ will be denoted by $\Omega_{m0}$,
$\Omega_{r0}$, and $\Omega_{\phi0}$, respectively.  $\Omega_{DE}$ will
denote
\begin{equation}
	\Omega_{DE} =
	\left\{ \begin{array}{ll}
	\Omega_\Lambda & \Lambda{\rm CDM} \\
	\Omega_\phi & {\rm QCDM~if~} w_\phi < -1/3 .\\
	\end{array} \right.
\end{equation}

Using WMAP5~\cite{Komatsu:2008hk} central values, we will set
$\Omega_{DE,0}$ = 0.72, $\Omega_{r0} = 8.5 \times 10^{-5}$,
$\Omega_{m0} = 1 - \Omega_{DE,0} - \Omega_{r0} \approx 0.28$, and
$\rho_{c0}^{1/4} = 2.5 \times 10^{-3}$ eV, with the present time $t_0
= 13.73$ Gyr after the big bang for $\Lambda$CDM.

The homogeneous scalar field obeys the Klein-Gordon equation
\begin{equation}
	\ddot{\phi} + 3 H \dot{\phi} = -\frac{d V}{d \phi} \equiv - V_\phi ~.
\label{phi}
\end{equation}
The evolution of the universe is described by the Friedmann equations
for the Hubble parameter $H = \dot{a}/a$ and the scale factor $a(t)$
\begin{equation}
	H^2 = \frac{\rho}{3 M_P^2} - \frac{k}{a^2}
\label{H}
\end{equation}
\begin{equation}
	\frac{\ddot{a}}{a} = - \frac{1}{6 M_P^2} (\rho + 3 P)
\label{acc}
\end{equation}
where the energy density $\rho = \rho_\phi + \rho_m + \rho_r$ and the
pressure $P = P_\phi + P_m + P_r$, with $P_m$ = 0, $P_r = \rho_r/3$,
and
\begin{equation}
	\rho_\phi = \frac{1}{2} \dot{\phi}^2 + V(\phi) ,~~
	P_\phi = \frac{1}{2} \dot{\phi}^2 - V(\phi) .
\label{rho-P}
\end{equation}
The curvature signature $k = +1$, 0, $-1$ for a closed, flat, or open
geometry.  Eq.~\eq{acc} shows that $P < -\rho/3$ for an accelerating
universe.

The conservation of energy equation for matter, radiation, and the
scalar field is
\begin{equation}
	\dot{\rho} + 3 H (\rho + P) = 0 .
\label{energy}
\end{equation}
Equation~\eq{energy} gives the evolution of $\rho_m$ and $\rho_r$, and
with Eq.~\eq{rho-P} the Klein-Gordon equation~\eq{phi} for the weakly
coupled scalar field.  The time rate of change of the Hubble parameter
is given by
\begin{equation}
	\dot{H} = - \frac{\rho + P}{2 M_P^2} + \frac{k}{a^2} ~.
\label{H-dot}
\end{equation}
Only two of Eqs.~\eq{H}, \eq{acc}, \eq{energy}, and~\eq{H-dot} are
independent.  We will assume a flat universe after inflation and
henceforth set $k$ = 0.

The logarithmic time variable (number of e-folds of the scale factor)
is defined as $\tau = \ln(a/a_0) = -\ln(1+z)$.  Note that for de
Sitter space $\tau = H_\Lambda t$, where $H_\Lambda^2 =
\rho_\Lambda/(3 M_P^2)$, and that $H_\Lambda t$ is a natural time
variable for the era of $\Lambda$-matter domination (see
e.g.\ Ref.~\cite{Gardner:2003nw}).  We will make the simple
approximations
\begin{equation}
	\rho_r = \rho_{r0} e^{-4\tau} ,~~
	\rho_m = \rho_{m0} e^{-3\tau} .
\end{equation}

The equation of state parameter for the scalar field $\phi$ is $w_\phi
= P_\phi/\rho_\phi$.  Since $\tau$ is a natural time variable for the
era of $\Lambda$-matter domination, we define the recent average of
$w_\phi$ as
\begin{equation}
	w_0 = \frac{1}{\tau} \int_0^z w_\phi d\tau .
\label{wbar}
\end{equation}
We will take the upper limit of integration to correspond to $z$ =
1.75.  The SNe Ia observations~\cite{Riess:2004nr} bound the recent
average $-1.1 < w_0 < -0.85$ (95\% CL).

For numerical simulations, the cosmological equations should be put
into a scaled, dimensionless form.  Equations~\eq{phi} and~\eq{H} can
be cast~\cite{Gardner:2004in} in the form of a system of two
first-order equations in $\tau$ plus a scaled version of $H$:
\begin{equation}
	\tilde{H} \frac{d \varphi}{d \tau} = \psi
\label{phi-tilde-1}
\end{equation}
\begin{equation}
	\tilde{H} \left(\frac{d \psi}{d \tau} + \psi\right) = 
	- 3 \tilde{V}_\varphi
\label{phi-tilde-2}
\end{equation}
\begin{equation}
	\tilde{H}^2 = \tilde{\rho} 
\label{scaled-H}
\end{equation}
\begin{equation}
	\tilde{\rho} = \frac{1}{6} \psi^2 + \tilde{V} + 
	\tilde{\rho}_m + \tilde{\rho}_r
\label{scaled-rho}
\end{equation}
where $\varphi \equiv \phi/M_P$, $\psi \equiv e^{2\tau}
\dot{\varphi}/H_0$, $\tilde{H} = e^{2\tau} H/H_0$, $\tilde{V} =
e^{4\tau} V/\rho_{c0}$, $\tilde{V}_\varphi = e^{4\tau}
V_\varphi/\rho_{c0}$, $\tilde{\rho} = e^{4\tau} \rho/\rho_{c0}$,
$\tilde{\rho}_m = e^{4\tau} \rho_m/\rho_{c0} = \Omega_{m0} e^\tau$,
$\tilde{\rho}_r = e^{4\tau} \rho_r/\rho_{c0} = \Omega_{r0}$.  This
scaling results in a set of equations that is numerically more robust,
especially near and before the time of big-bang nucleosynthesis
(BBN)---see Ref.~\cite{Gardner:2004in}, especially Fig.~1.

For a contracting phase (in which $H$ goes through zero), a different
set of equations and a different scaling should be used.  Here we will
use the conformal time variable
\begin{equation}
	\eta = \int_0^t \frac{a_0 H_0}{a} dt 
\label{eta}
\end{equation}
where $t = 0$ corresponds to the big bang.

Equations~\eq{phi} and~\eq{H-dot} can be cast in the form of a system of
three first-order equations in $\eta$:
\begin{equation}
	\frac{d \varphi}{d \eta} = \psi
\label{phi-bar-1}
\end{equation}
\begin{equation}
	\frac{d \psi}{d \eta} = -2 \overline{H} \psi - 3 \overline{V}_\varphi
\label{phi-bar-2}
\end{equation}
\begin{equation}
	\frac{d \overline{H}}{d \eta} =
	- \frac{1}{2} \left(\overline{\rho} + 3 \overline{P}\right) =
	- \frac{1}{3} \psi^2 + \overline{V} - \frac{\Omega_{m0}}{2} e^{-\tau} -
	\Omega_{r0} e^{-2\tau}
\label{H-bar}
\end{equation}
\begin{equation}
	\overline{\rho} = \frac{1}{6} \psi^2 + \overline{V} + 
	\overline{\rho}_m + \overline{\rho}_r ,~~
	\overline{P} = \frac{1}{6} \psi^2 - \overline{V} + 
	\frac{1}{3} \overline{\rho}_r
\label{P-rho}
\end{equation}
where $\varphi \equiv \phi/M_P$, $\psi \equiv e^{\tau}
\dot{\varphi}/H_0$, $\overline{H} = e^{\tau} H/H_0$, $\overline{V} =
e^{2\tau} V/\rho_{c0}$, $\overline{V}_\varphi = e^{2\tau}
V_\varphi/\rho_{c0}$, $\overline{\rho} = e^{2\tau} \rho/\rho_{c0}$,
$\overline{\rho}_m = e^{2\tau} \rho_m/\rho_{c0} = \Omega_{m0}
e^{-\tau}$, $\overline{\rho}_r = e^{2\tau} \rho_r/\rho_{c0} =
e^{-2\tau} \Omega_{r0}$.  This scaling results in a set of numerically
more robust equations, especially near the turn-around time $t_*$
between expanding and contracting phases of the universe.

Note that the conformal time $\eta$ is related to the logarithmic time
$\tau$ by
\begin{equation}
	\frac{d \tau}{d \eta} = \overline{H} .
\label{dtau}
\end{equation}

\section{Simulations of Unstable Axion Quintessence}

The original axion quintessence potential $V = A (1+\cos(\varphi))$
was based on $N = 1$ supergravity~\cite{Frieman:1995pm,Waga:2000ay},
with $m_\phi^2 = 3 H_\Lambda^2$.  As $\varphi \rightarrow \pi$, the
universe evolves to Minkowski space.

The unstable de Sitter axion potential $V = A \cos(\varphi)$ is based
on M/string theory reduced to an effective $N = 1$ supergravity
theory~\cite{Choi:1999xn}, with $m_\phi^2 = -3 H_\Lambda^2$ at the
maximum of $V$.

Both axion quintessence models are derivable (up to a constant) from
string theory as axion monodromy~\cite{Silverstein:2008sg}.

The quintessence axion is a pseudo Nambu-Goldstone boson: at the
perturbative level the theory is shift symmetric under $\varphi
\rightarrow \varphi + const$ with $\varphi = \phi/M$.  The shift
symmetry is broken---before or during inflation---by nonperturbative
instanton effects to a discrete symmetry $\varphi \rightarrow \varphi
+ 2 \pi$, generating a potential $V(\varphi) = A (C + \cos(\varphi))$.
In these theories, quantum corrections to the classical axion
potential are suppressed.  For quintessence (or for natural
inflation~\cite{Freese:2008if}), $M \sim M_P$; we will take $M = M_P$.
$C = 0$ and $C = 1$ are the most interesting unstable axion and
original axion cases, respectively.

For the unstable axion quintessence computations (with an expanding
and contracting universe), we will use Eqs.~\eq{phi-bar-1}--\eq{H-bar}
and~\eq{dtau} with initial conditions $\varphi_i$ and $\dot{\varphi}_i
= 0$ specified at matter-radiation equality $z_{mr} = 3280$, which
corresponds to
\begin{equation}
	\eta_{mr} = \frac{2\left(\sqrt{2}-1\right)}
        {\sqrt{\Omega_{m0}} \sqrt{1+z_{mr}}} ~.
\end{equation}
The constant $A$ in the potential is adjusted so that $\Omega_{\phi0}$
= 0.72.  This involves the usual single fine tuning.  Note that the
same anthropic arguments that limit the magnitude of a present-day
cosmological constant also limit $A < 100 \rho_{c0}$, so the tuning of
$A$ is no worse than the tuning of a cosmological constant.

\begin{table}[htb]
\center{
\begin{tabular}{|c c c c c c c c c c|} \hline 
$\varphi_i/\pi$ & $A/\rho_{c0}$ & $w_0$ & $t_0$ & $t_{0.1}$ & $t_{0.9}$ & 
  $t_*$ & $t_f$ & $\Delta t_c/t_f$ & $a_*/a_0$ \\ \hline
$0.05$ & 0.73 & $-0.998$ & 13.72 & 3.6 & 20.0     & 63.2 & 72.7 & 0.23 & 12.0\\
$0.10$ & 0.78 & $-0.99$  & 13.70 & 3.5 & 20.3     & 47.6 & 56.8 & 0.30 & 5.0 \\
$0.15$ & 0.88 & $-0.97$  & 13.64 & 3.5 & 21.2     & 37.6 & 46.2 & 0.38 & 3.0 \\
$0.20$ & 1.09 & $-0.93$  & 13.49 & 3.3 & 21.2$^*$ & 29.2 & 37.0 & 0.48 & 2.0 \\
$0.23$ & 1.41 & $-0.87$  & 13.25 & 3.0 & 16.8$^*$ & 23.9 & 30.7 & 0.45 & 1.6\\
\hline
\end{tabular}
}
\caption{Parameters for the potential $V = A \cos(\varphi)$.  $t_0$ is
  the current age of the universe with $t_0 \equiv 13.73$ Gyr in the
  $\Lambda$CDM model, $0.1 \le \Omega_{DE} \le 0.9$ for $t_{0.1} \le t
  \le t_{0.9}$, the ``coincidence'' time interval $\Delta t_c =
  t_{0.9}-t_{0.1}$, $t_*$ is the turn-around time, $t_f$ is the time
  of the big crunch, and $a_* = a(t_*)$.  All times are in Gyr.
  $^*$For $\varphi_i/\pi$ = 0.20 (0.23), $\Omega_{DE} \le$ 0.85 (0.77)
  and in these cases $t_{0.9} \equiv t_{0.85}$ and $t_{0.77}$,
  respectively.}
\label{cos}
\end{table}

\begin{figure}[htbp]
\begin{center}
\scalebox{1.05}{\includegraphics{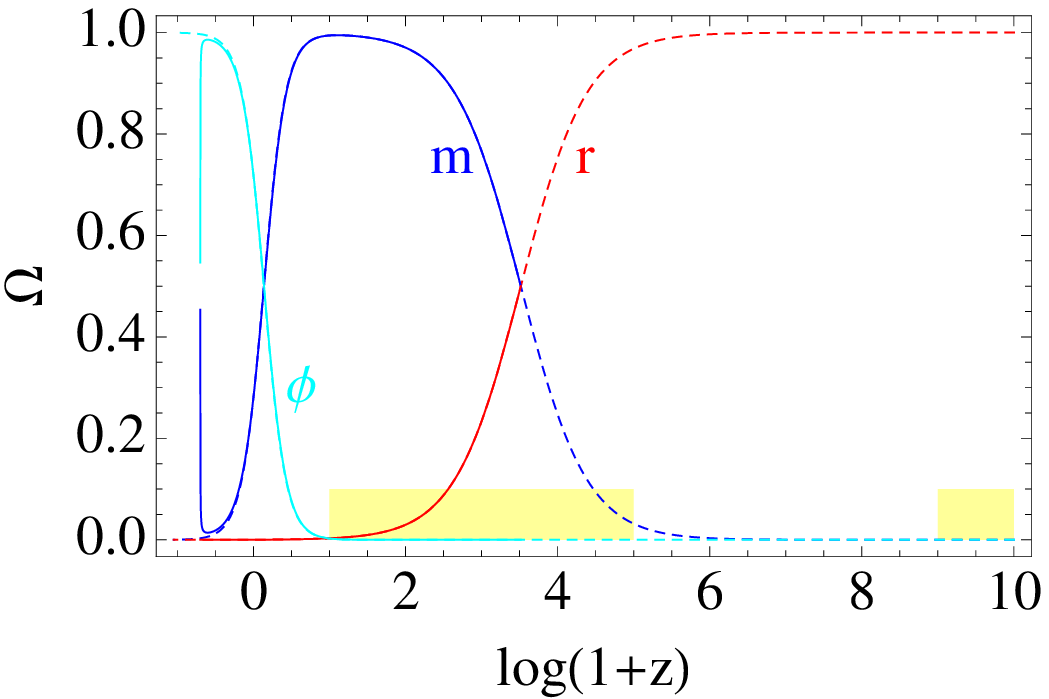}}
\end{center}
\caption{$\Omega$ vs.\ $\log_{10}(1+z)$ for the potential $V = A
  \cos(\varphi)$, $\varphi_i/\pi$ = 0.1 (solid) vs.\ $\Lambda$CDM
  (dotted).  The light yellow rectangles are the bounds on
  $\Omega_{DE}$ from LSS, CMB, and BBN.}
\label{fig-Omega}
\end{figure}

\begin{figure}[htbp]
\begin{center}
\scalebox{1.05}{\includegraphics{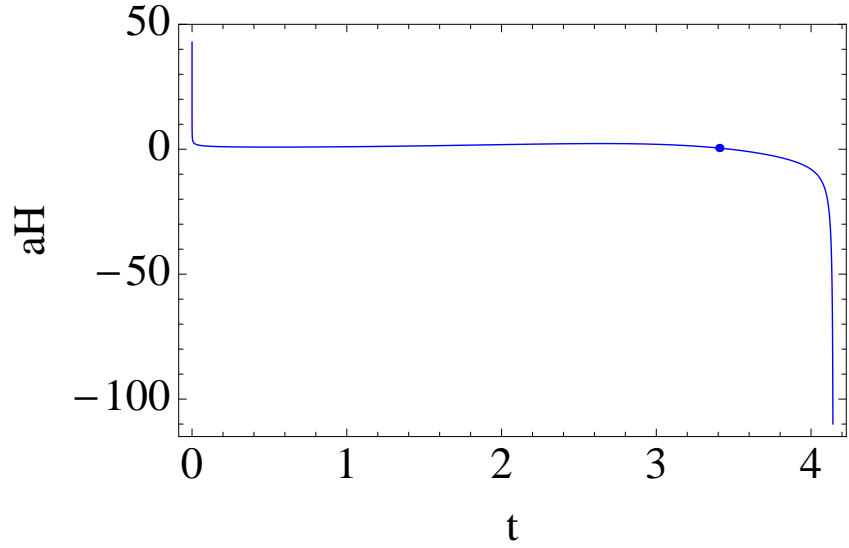}}
\caption{Comoving Hubble parameter $a H/(a_0 H_0)$ vs.\ $t/t_0$.  The
  dot indicates the value $H = 0$ at $t_*$.}
\label{fig-aH}
\end{center}
\end{figure}

\begin{figure}[htbp]
\begin{center}
\scalebox{1.05}{\includegraphics{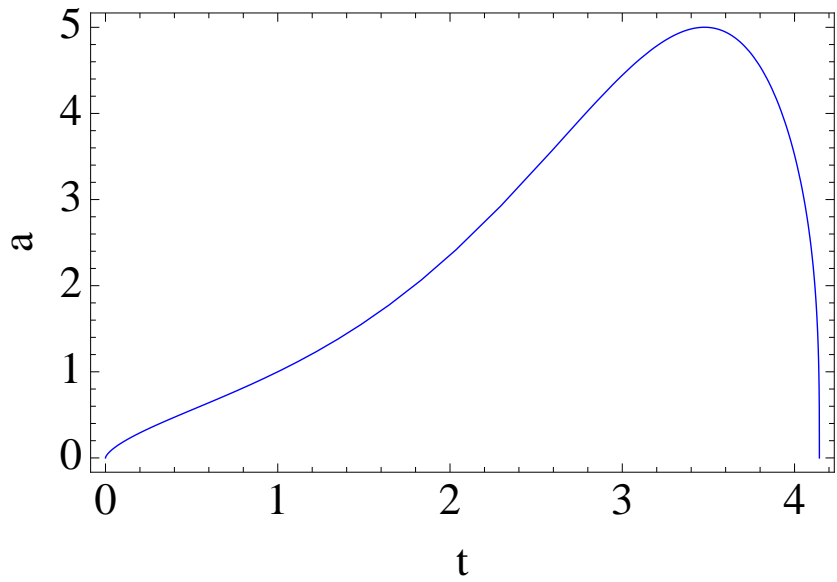}}
\caption{Scale factor $a/a_0$ vs.\ $t/t_0$.}
\label{fig-a}
\end{center}
\end{figure}

\begin{figure}[htbp]
\begin{center}
\scalebox{1.05}{\includegraphics{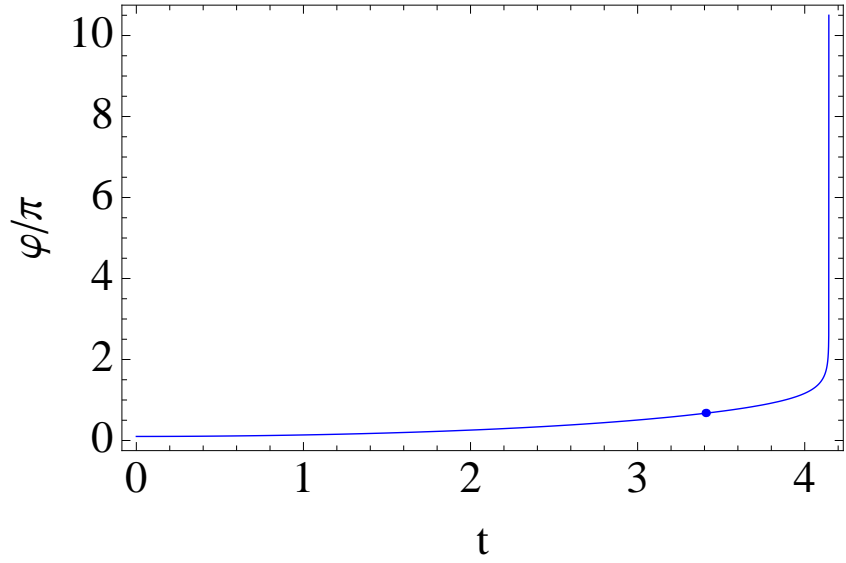}}
\caption{$\varphi/\pi$ vs.\ $t/t_0$.  The dot indicates the value
  $\varphi/\pi = 0.67$ at $t_*$.}
\label{fig-phi}
\end{center}
\end{figure}

\begin{figure}[htbp]
\begin{center}
\scalebox{1.05}{\includegraphics{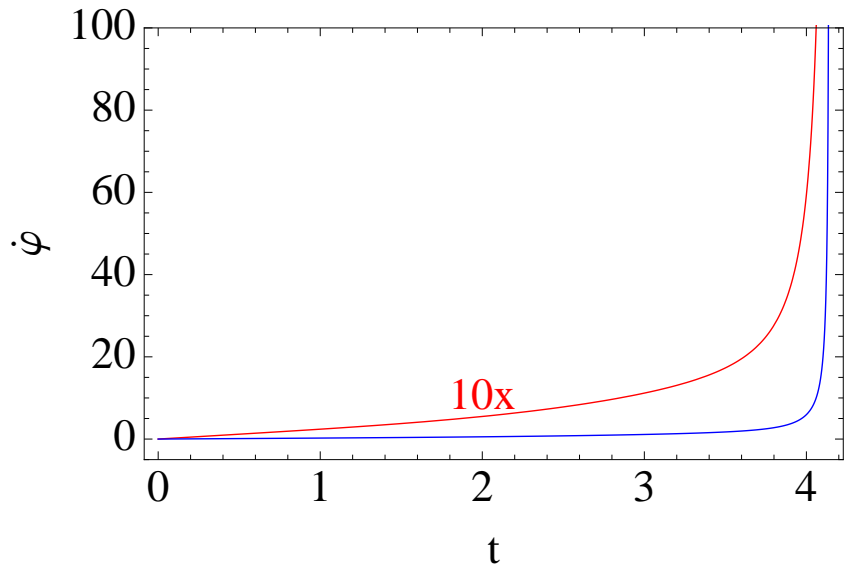}}
\caption{$\dot{\varphi}/H_0$ and $10\, \dot{\varphi}/H_0$ vs.\ $t/t_0$.}
\label{fig-psi}
\end{center}
\end{figure}

\begin{figure}[htbp]
\begin{center}
\scalebox{1.05}{\includegraphics{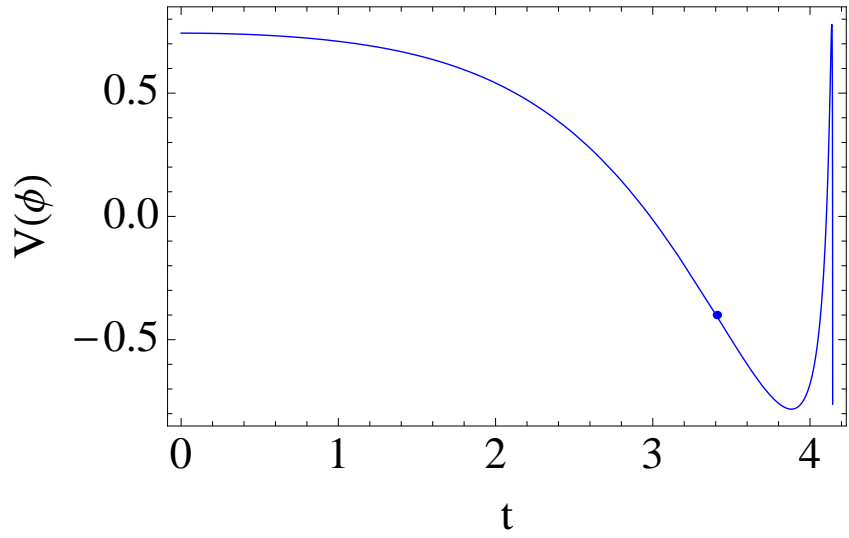}}
\caption{Potential $V(\phi)/\rho_{c0}$ vs.\ $t/t_0$.  The dot
  indicates the value of $V/\rho_{c0}$ at $t_*$.}
\label{fig-V}
\end{center}
\end{figure}

\begin{figure}[htbp]
\begin{center}
\scalebox{1.05}{\includegraphics{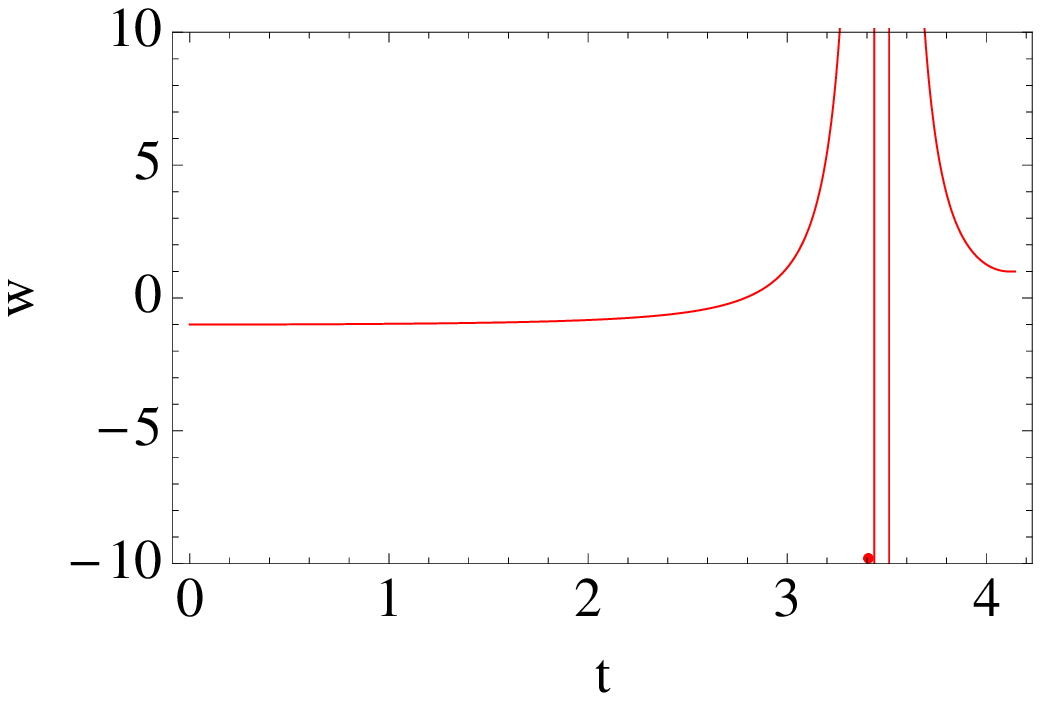}}
\end{center}
\caption{$w_\phi$ vs.\ $t/t_0$.  The dot indicates $t_*$.}
\label{fig-w}
\end{figure}

\begin{figure}[htbp]
\begin{center}
\scalebox{1.05}{\includegraphics{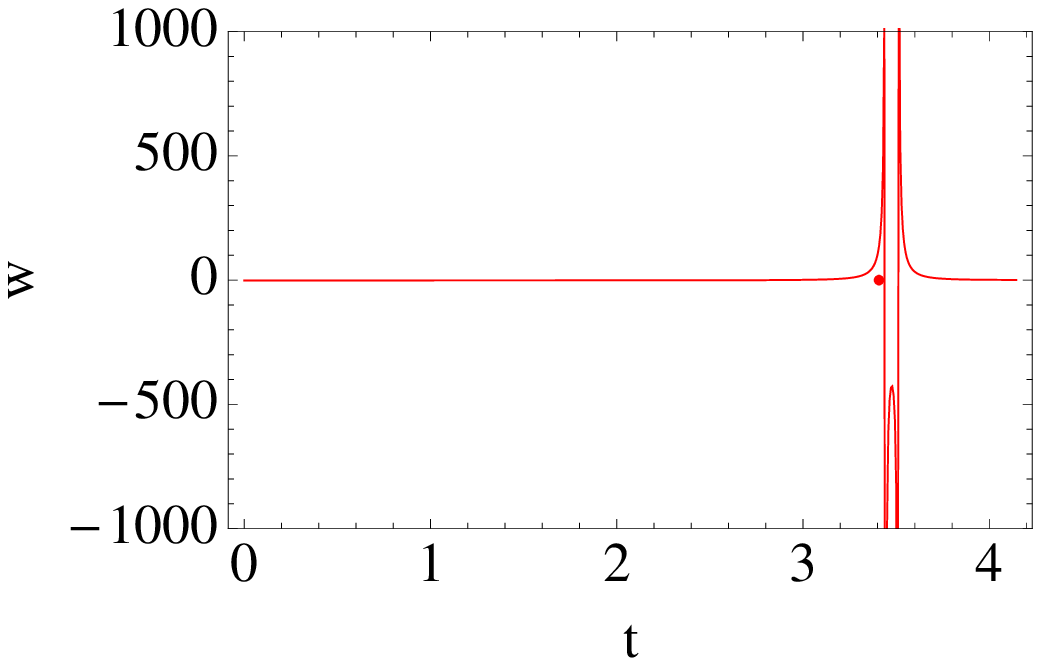}}
\end{center}
\caption{Overview of $w_\phi$ vs.\ $t/t_0$.  The dot indicates $t_*$.}
\label{fig-w-o}
\end{figure}

\begin{figure}[htbp]
\begin{center}
\scalebox{1.05}{\includegraphics{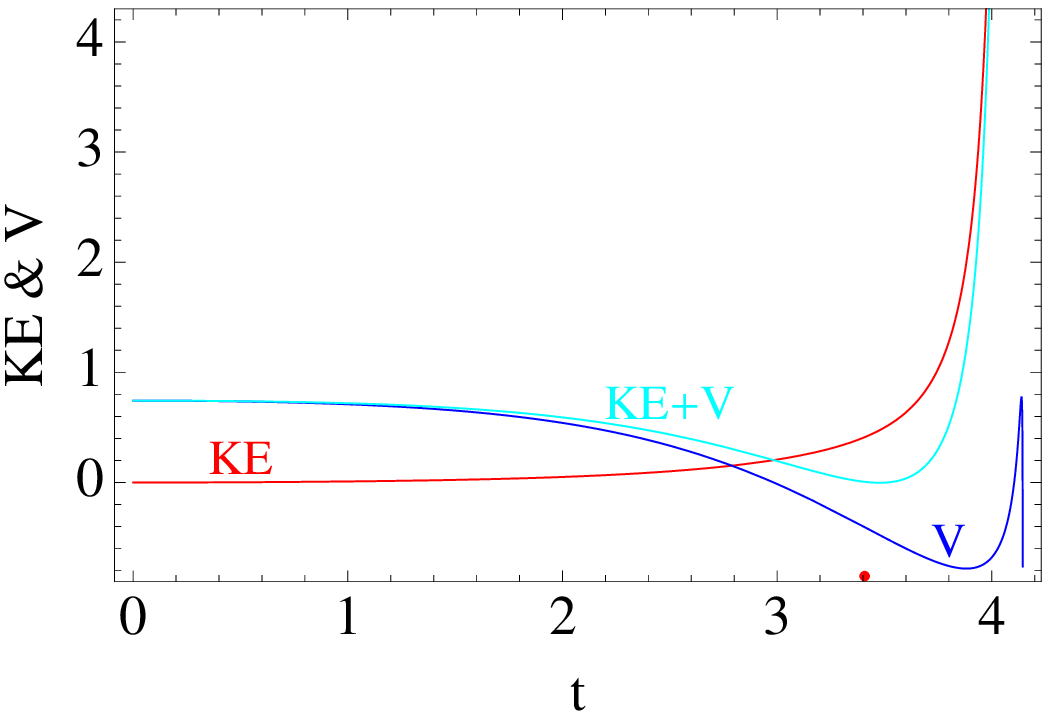}}
\end{center}
\caption{Scalar field kinetic energy density $\dot{\phi}^2/2$ and
  potential energy density $V(\phi)$ vs.\ $t/t_0$.  The dot
  indicates $t_*$.}
\label{fig-KE-V}
\end{figure}

Results for the unstable axion potential are presented in
Table~\ref{cos} for various $\varphi_i$ and for $\varphi_i/\pi = 0.1$
in Figs.~\ref{fig-Omega}--\ref{fig-KE-V}.  (As $\varphi_i \rightarrow
0$, classically $t_f \rightarrow \infty$, but quantum effects
destabilize $\varphi_i \approx 0$ so that the maximum $t_f \sim 100\,
t_0$~\cite{Kallosh:2002gf}.)  For the values in the Table, as
$\varphi_i$ increases, $\left|V_\phi\left(\phi_i\right)\right|$ also
increases and $\phi$ starts to move earlier, leading to a decrease in
$t_{0.1}$, $t_0$, $t_*$, and $t_f$, and correspondingly to an increase
in $w_0$ away from $-1$.  Note that for $\varphi_i/\pi = 0.2$, the
coincidence time ratio approaches 50\%.

The QCDM universe mimics the $\Lambda$CDM model (see
Fig.~\ref{fig-Omega}; for clarity, only the beginnings of the
contracting stage are shown in this figure) until about $z = -0.5$,
after which the QCDM universe begins to decelerate and ultimately to
rapidly contract to a big crunch (Figs.~\ref{fig-aH} and~\ref{fig-a}).

During the contracting stage, $H < 0$ acts as a {\em negative}\/
friction in
\begin{displaymath}
	\ddot{\phi} + 3 H \dot{\phi} + V_\phi = 0
\end{displaymath}
amplifying the axion field and its kinetic energy to bring about a
late stage kination era during which the scalar field kinetic energy
dominates over all other forms of energy.

The quintessence axion is an ultra-light scalar field with $m_\phi^2
\sim H_\Lambda^2$, so $\phi$ ``sits and waits'' during the early
evolution of the universe, and only starts to move when $H^2 \sim
m_\phi^2$ (Figs.~\ref{fig-phi} and~\ref{fig-psi}).  In this way it is
easy to satisfy the BBN ($z \sim 10^9$--$10^{11}$), cosmic microwave
background (CMB) ($z \sim 10^3$--$10^5$), and large scale structure
(LSS) ($z \sim 10$--$10^4$) bounds on $\Omega_{DE} \<~ 0.1$, as in
Fig.~\ref{fig-Omega}.  An ultra-light scalar field also reflects the
observational evidence that the universe has only recently become
dominated by dark energy.

In Fig.~\ref{fig-aH}, the Hubble parameter goes through zero at the
turn-around time between an expanding and contracting universe.  At
the beginning of the contacting stage, $\phi$ has yet to reach the
minimum of the potential energy (see Fig.~\ref{fig-V}), and thus the
negative Hubble parameter amplifies the kinetic energy of the scalar
field, bringing about an era of kination with $w_\phi = 1$, as seen in
Figs.~\ref{fig-phi}, \ref{fig-psi}, \ref{fig-w}, and~\ref{fig-w-o}.
Also note that Figs.~\ref{fig-aH}, \ref{fig-phi}, and~\ref{fig-psi}
indicate that $H$, $\phi$, and $\dot{\phi}$ are approaching a
singularity near $t_f$.  In fact, in an era of kination during
contraction during which $H = -\left|\dot{\phi}\right|/(\sqrt{6}
M_P)$, $\dot{\varphi} = \sqrt{2/3}/\left(t_f-t\right)$ and $\varphi =
-\sqrt{2/3}\ln\left(t_f-t\right)$, while $a \sim
\left(t_f-t\right)^{1/3}$ (see Fig.~\ref{fig-a}).

Figure~\ref{fig-V} shows that as $\phi$ increases without bound, the
potential energy $V(\phi)$, since a periodic function of $\phi$,
oscillates more and more rapidly.  Depending on the strength of the
coupling between the quintessence axion and other particles, $\phi$
may decay and populate the universe with additional radiation and
matter.  (The monotonically increasing field $\phi$ in a periodic
potential can be interpreted as an oscillating field.)  At the very
least, there should be gravitational production of particles by $\phi$
during contraction.

Figures~\ref{fig-w} and~\ref{fig-w-o} follow the contracting stage
further, illustrating that although $w_\phi \rightarrow \pm \infty$
twice just after $t_*$, the universe ultimately enters a stage of
kination in which $w_\phi = 1$ near $t_f$.  Note that $w_\phi \approx
-1$ until well after $t_0$.

After matter-quintessence equality at $t_{m\phi} \approx 0.7\, t_0$,
the scalar field energy density always dominates over the matter and
radiation energy densities.  There is a period from $t \approx 2.5\,
t_0$ until $3.8\, t_0$ when the scalar field potential energy is
comparable to its kinetic energy, and then the kinetic energy (which
scales as $1/a^6$) predominates during the rapid contraction to a big
crunch (see Fig.~\ref{fig-KE-V}).

\section{Conclusion}

As the universe contracts, density inhomogeneities are amplified and
presumably black holes are formed, similarly to the contracting stage
of the ekpyrotic universe~\cite{Lehners:2009eg,Khoury:2001zk} with $w
= 1$.  Depending on the strength of the coupling (which we have
neglected here) between the quintessence axion and other particles,
$\phi$ may decay and produce radiation and matter.  There should at
least be gravitational production of particles by $\phi$ during
contraction.  As the universe reheats during contraction, broken
symmetries are restored.  It is possible that inflating patches may be
generated, spawning new universes from the old.  These issues are
currently under investigation.

In summary, the unstable axion quintessence potential resolves the
issues concerning quintessence raised by Vilenkin: the minimum of the
potential is not at zero, but at a negative value $\approx
-\rho_\Lambda$, $\rho_{DE,0} = \rho_\Lambda$ with only a single fine
tuning (in the anthropic range), and $w_0$ naturally satisfies $-1 <
w_0 \le -0.87$, for an appreciable 23\% range of possible initial
values for the quintessence field.  And for a universe like ours, the
coincidence time when the energy densities of dark energy and matter
are comparable varies (as long as $\phi_i/\pi$ is not too small---say,
$\ge$ 0.05) from 25\%--50\% of the lifetime of the universe.

\section*{Acknowledgement}

My thanks are due to Lawrence Krauss for valuable comments.

\newpage

\begin{thebibliography}{10}

\bibitem{Kallosh:2002gf}
  Kallosh~R, Linde~A~D, Prokushkin~S, and Shmakova~M,
  {\em Supergravity, dark energy and the fate of the universe,}
  2002 Phys.\ Rev.\  D {\bf 66} 123503 
  [arXiv:hep-th/0208156].

\bibitem{Krauss:1999br}
  Krauss~L~M and Turner~M~S,
  {\em Geometry and destiny,}
  1999 Gen.\ Rel.\ Grav.\  {\bf 31} 1453 
  [arXiv:astro-ph/9904020].

\bibitem{Vilenkin:2009zz}
  Vilenkin~A,
  {\em Perspectives in cosmology,}
  2010 J.\ Phys.\ Conf.\ Ser.\  {\bf 203} 012001 
  [arXiv:0908.0721 [astro-ph.CO]].

\bibitem{Gardner:2004in}
  Gardner~C~L,
  {\em Quintessence and the transition to an accelerating universe,}
  2005 Nucl.\ Phys.\  B {\bf 707} 278 
  [arXiv:astro-ph/0407604].

\bibitem{Komatsu:2008hk}
  Komatsu~E {\it et al.}  [WMAP Collaboration],
  {\em Five-year Wilkinson Microwave Anisotropy Probe (WMAP)
  observations: Cosmological interpretation,}
  2009 Astrophys.\ J.\ Suppl.\  {\bf 180} 330 
  [arXiv:0803.0547 [astro-ph]].

\bibitem{Gardner:2003nw}
  Gardner~C~L,
  {\em Cosmological variation of the fine structure constant from an 
  ultra-light scalar field: The effects of mass,}
  2003 Phys.\ Rev.\  D {\bf 68} 043513 
  [arXiv:astro-ph/0305080].

\bibitem{Riess:2004nr}
  Riess~A~G {\it et al.}  [Supernova Search Team Collaboration],
  {\em Type Ia supernova discoveries at z>1 from the Hubble Space Telescope:
  Evidence for past deceleration and constraints on dark energy evolution,}
  2004 Astrophys.\ J.\  {\bf 607} 665 
  [arXiv:astro-ph/0402512].

\bibitem{Frieman:1995pm}
  Frieman~J~A, Hill~C~T, Stebbins~A, and Waga~I,
  {\em Cosmology with ultralight pseudo Nambu-Goldstone bosons,}
  1995 Phys.\ Rev.\ Lett.\  {\bf 75} 2077 
  [arXiv:astro-ph/9505060].

\bibitem{Waga:2000ay}
  Waga~I and Frieman~J~A,
  {\em New constraints from high redshift supernovae and lensing statistics 
  upon scalar field cosmologies,}
  2000 Phys.\ Rev.\  D {\bf 62} 043521 
  [arXiv:astro-ph/0001354].

\bibitem{Choi:1999xn}
  Choi~K,
  {\em String or M theory axion as a quintessence,}
  2000 Phys.\ Rev.\  D {\bf 62} 043509 
  [arXiv:hep-ph/9902292].

\bibitem{Silverstein:2008sg}
  Silverstein~E and Westphal~A,
  {\em Monodromy in the CMB: Gravity waves and string inflation,}
  2008 Phys.\ Rev.\  D {\bf 78} 106003 
  [arXiv:0803.3085 [hep-th]].

\bibitem{Freese:2008if}
  Freese~K, Savage~C, and Kinney~W~H,
  {\em Natural Inflation: the status after WMAP 3-year data,}
  2008 Int.\ J.\ Mod.\ Phys.\  D {\bf 16} 2573 
  [arXiv:0802.0227 [hep-ph]].

\bibitem{Lehners:2009eg}
  Lehners~J~L, Steinhardt~P~J, and Turok~N,
  {\em The return of the Phoenix Universe,}
  2009 Int.\ J.\ Mod.\ Phys.\  D {\bf 18} 2231 
  [arXiv:0910.0834 [hep-th]].

\bibitem{Khoury:2001zk}
  Khoury~J, Ovrut~B~A, Steinhardt~P~J, and Turok~N,
  {\em Density perturbations in the ekpyrotic scenario,}
  2002 Phys.\ Rev.\  D {\bf 66} 046005 
  [arXiv:hep-th/0109050].

\end{thebibliography}
\end{document}